\documentclass[a4paper]{jpconf} 

\usepackage{eurosym} 
\usepackage{amsmath} 
\usepackage{amssymb} 
\usepackage{dsfont} 
\usepackage{enumerate} 

\usepackage{graphicx} 

\usepackage[mathscr]{eucal} 
\usepackage{longtable} 
\usepackage{multirow}

\usepackage{color}

\newcommand{\dd}{\mathrm{d}}
\newcommand{\Dp}{\partial}

\newcommand{\ve}{\varepsilon}

\newcommand{\li}{\left}
\newcommand{\ri}{\right}

\newcommand{\cen}[1]{\begin{center} #1 \end{center}}

\newcommand{\tc}{T_{{\rm c}}}

\usepackage{lscape}
\usepackage{rotating}

\begin{document}
\title{Holography at QCD-$\boldsymbol \tc$}

\author{R. Z\"ollner, B. K\"ampfer}
\address{Helmholtz-Zentrum Dresden-Rossendorf, PF 510119, D-01314 Dresden, Germany \\ and \\ Institut f\"ur Theoretische Physik, TU Dresden, D-01062 Dresden, Germany}

\begin{abstract}
Within an extended soft wall model, we study the temperature dependence of the lowest vector meson states. Scales are adjusted by using as input the $\rho$ meson mass in vacuum and the velocity of sound from lattice QCD which displays a minimum at $\tc$. The melting of the $\rho$ meson occurs at temperatures $\mathcal{O}(\tc)$. 
\end{abstract}

\section{Introduction}
The soft wall model \cite{KKSS} sparked a series of applications by describing hadrons within the AdS/CFT correspondence. Interesting are, for instance, extensions to non-zero temperatures as initiated in \cite{Col09,Col12,x1}. A particular formulation of the soft wall has been put forward in \cite{ich3}, based on \cite{ich1}, where some relations to QCD have been considered on a qualitative level. Simple parametrizations have been proposed to accommodate QCD thermodynamics features together with a proper temperature dependence of the light-quark vector meson spectrum (for non-zero net baryon effects, cf.~\cite{ich2}). Here, we explore quantitatively the compatibility of the setting in \cite{ich3} with (i) QCD thermodynamics, in particular the velocity of sound (Section 2) and (ii) the melting of the $\rho$ meson (Section 3) within the extended soft wall model (\ref{appa}). Our conclusion (Section 4) is that a semi-quantitative compatibility of items (i) and (ii) can be accomplished which leaves space for further fine tuning (\ref{appb}).

\section{Thermodynamics}
In a self consistent `dynamical` model, e.g.~the setting pioneered in \cite{Gubser} (cf.~\cite{Kir1, Kir2} for a complementing approach) and further developed in \cite{Finazzo1,Knaute} to accommodate the case of 2+1 flavor QCD thermodynamics with physical quark masses, the ansatz for the metric of a 5dimensional Riemann space
  \begin{equation}
   \dd s^2 = e^{A(z)} \li(f(z) \dd t^2 -\dd \vec x^{\,2} - \frac{\dd z^2}{f(z)}\ri) \label{ds}
  \end{equation}
is employed to extract the metric functions $A(z)$, $f(z)$ as well as the dilaton profile $\Phi(z)$ by solving the Einstein equations and equation of motion resulting from the action
  \begin{equation}
   S = \frac1{16 \pi G_5} \int \! \dd^4x \, \dd z \sqrt{g} \li(R-\frac12 (\Dp \Phi)^2 -\tilde V(\Phi) \ri). \label{dilatonwirkung}
  \end{equation}
Instead of tuning the dilaton potential $\tilde V(\Phi)$ appropriately, as done in \cite{Knaute} for instance (cf.~also \cite{Finazzo1,fn1,fn2,fn3}; for the Yang-Mills case, cf.~\cite{Yaresko}), we postpone that step for follow-up investigations and consider suitable ans\"atze. Such ans\"atze must be conform with the AdS behavior near the boundary at $z=0$ and, for non-zero temperatures, must display a simple zero of the blackness function $f(z)$ at the horizon $z_H$. As demonstrated in \cite{ich3} the parametrizations of Hawking temperature $T$ and Bekenstein entropy density $s$
  \begin{eqnarray}
   T(z_H) &=& \frac{T_x}{\theta} \li(\frac1{ x} + \theta - {3b}+ {3bx} - {x^2} \ri), \label{T} \\
   s(z_H) &=& \frac{16\pi G_5}{z_H^3} \label{s}
  \end{eqnarray}
with $x=z_H/z_x$ and parameters $T_x$, $\theta=\pi z_x T_x$ and $b$ generate a phase structure in agreement with 2+1 flavor QCD with varying quark masses: $b>1$ delivers a first-order phase transition, $b=1$ is for a second-order phase transition and $b<1$ implies a cross over.\footnote{An inflection point of $T(z_H)$ in conjunction with (\ref{s}) is a sufficient condition for a local minimum of $c_s^2$. In fact, (\ref{T}) has such an inflection at $x=1$ and $T(x=1) = T_x$. The example $T(z_H)=1/(\pi z_H) +T_x -T_y^2z_H/\pi$ proves that an inflection point is not necessary to generate a minimum of $c_s^2(T)$. We use here, however, (\ref{T}) as in \cite{ich3}.} 
This scans through the Columbia plot \cite{Philipsen:2015eya} on the line of degenerate quark masses. Here, we restrict ourselves to $b<1$ and compare with the lattice QCD results \cite{bor14,baza14}. The velocity of sound
  \begin{equation}
   c_s^2 = \frac{s}{T} \frac{\dd T}{\dd s}
  \end{equation}
is a preferred quantity since it requires a minimum of scale settings. 

  \begin{figure}
   \cen{\includegraphics[scale=.7]{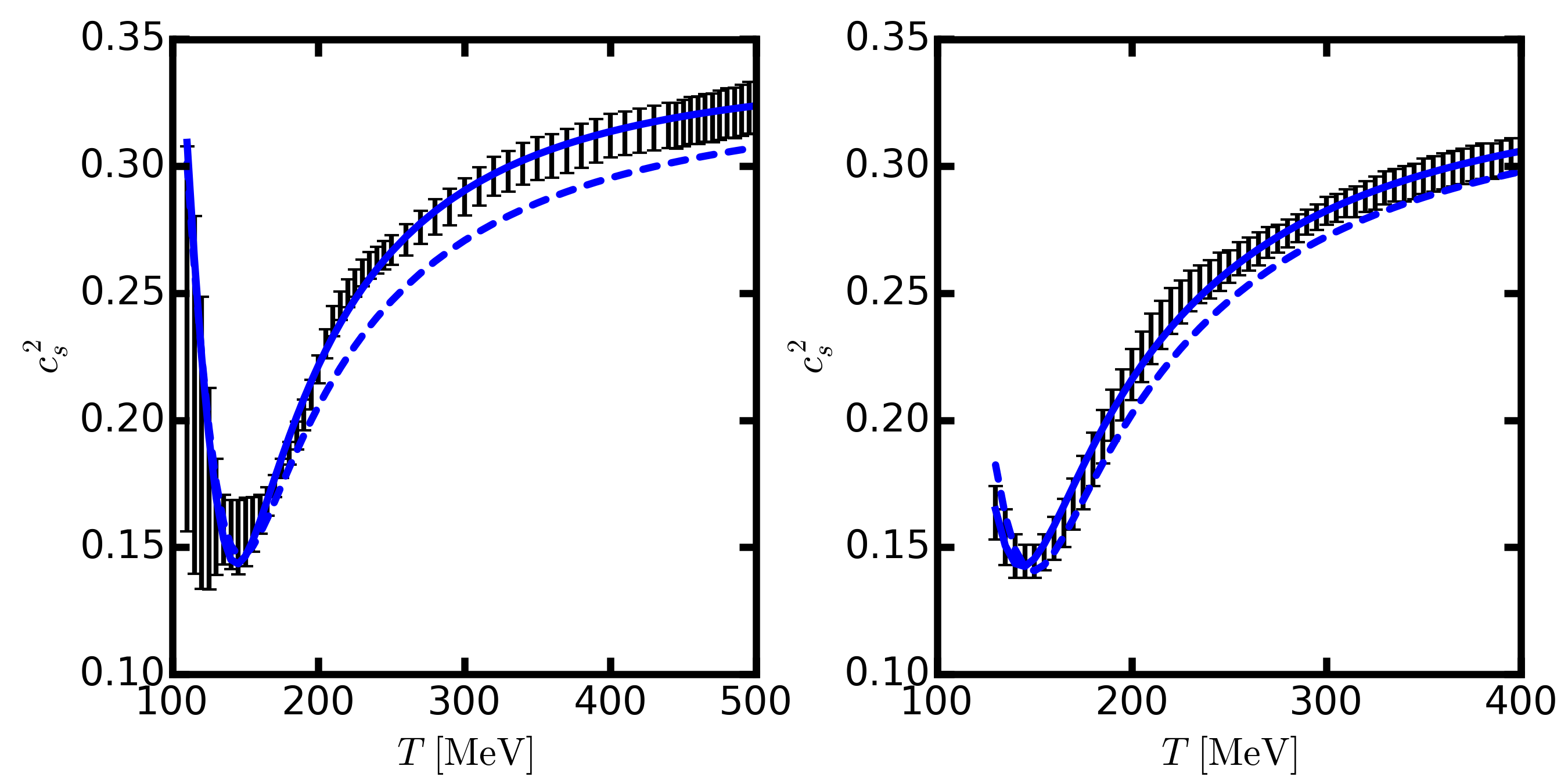}
   \caption{Speed of sound squared as a function of the temperature. Data from \cite{bor14} (left panel) and \cite{baza14} (right panel) are for the 1$\sigma$ error range. Solid curves: minimization of $\chi^2$ delivering $T_x=$130 (129) MeV, $\theta=$1.75 (1.90) and $b=$0.71 (0.68) for the left (right) panel. Dashed curves: model adjustment to the minimum with coordinates $\tc=$146 (150) MeV and $c_s^2(\tc)=0.147$ (0.14) and the side condition $T_{\rm dis}^{\rho}=$145 (150) MeV; parameters are $T_x=$129 (129) MeV, $\theta=$2.07 (2.02) and $b=$0.62 (0.66) for the left (right) panel.}\label{c2_abb2}}
  \end{figure}
  
\begin{figure}
   \cen{\includegraphics[scale=.7]{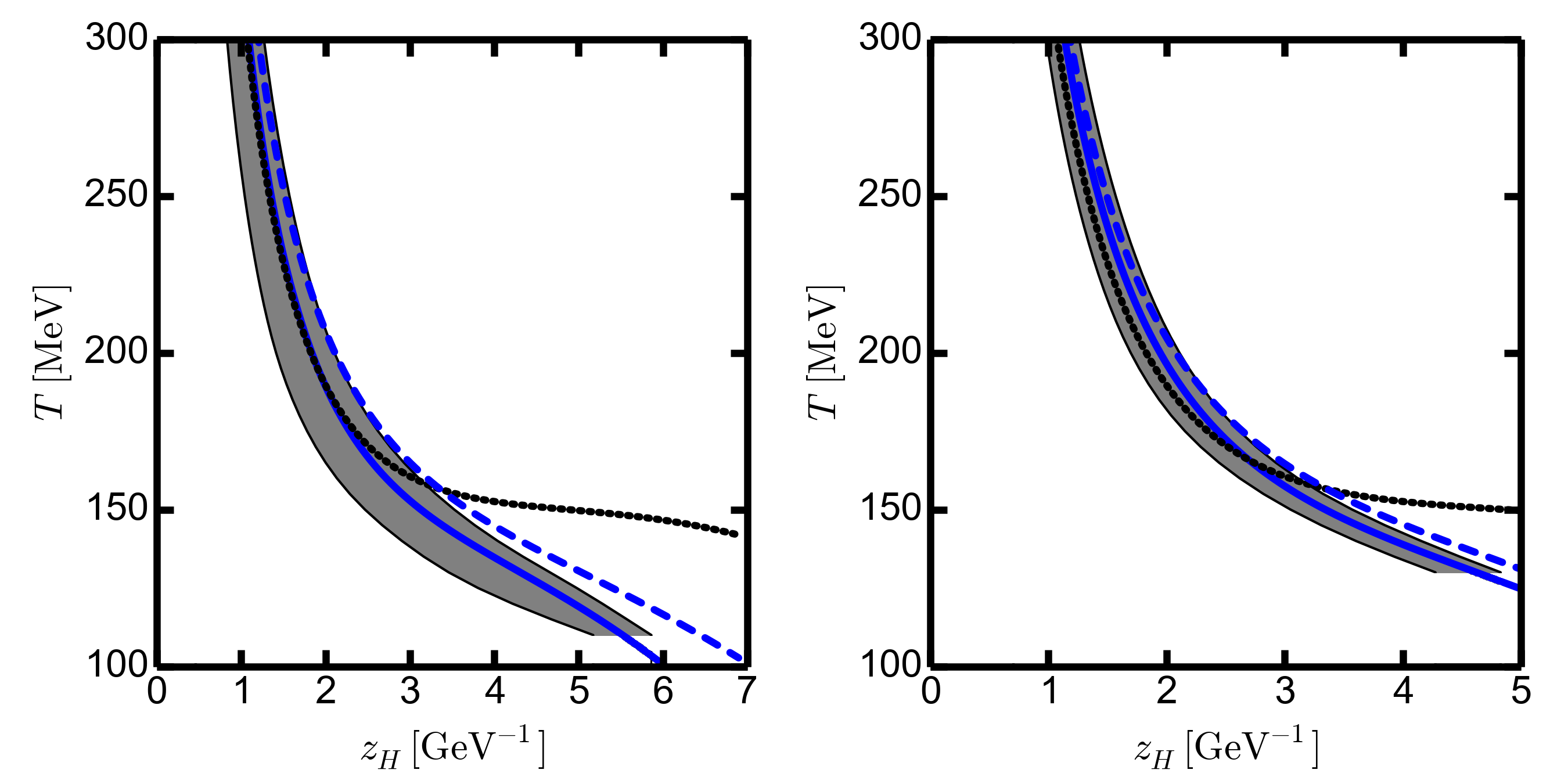}
   \caption{The temperature as a function of the horizon position $z_H$ obtained by (\ref{zh}) with the data displayed in Fig.~\ref{c2_abb2} (gray band). The solid and dashed curves correspond to those in Fig.~\ref{c2_abb2}. The dotted curve belongs to the parameter set $T_x=150$ MeV, $\theta=2.89$ and $b=0.94$ which accomplishes $T_{\rm dis}^{\rho}=155$ MeV.}\label{c2_abb3}}
  \end{figure}

Figure \ref{c2_abb2} exhibits two fits of the sound velocity squared, separately for the data \cite{bor14} (left) and \cite{baza14} (right). The solid curve is for a minimized $\chi^2$, while the dashed curve is adjusted at the minimum of $c_s^2$ (softest point) fixed by the parameters $c_s^2(\tc)=\min\{c_s^2\}=0.147$ (left, 0.14 right) at the temperature $\tc=146$~MeV (left, 150~MeV right) and a $\rho$ meson melting (disappearance) temperature of $T_{\rm dis}^{\rho}=145$~MeV (left, 150~MeV right).\footnote{The equation of state has several distinctive points: the minimum of sound velocity at $\tc^{c_s^2} = 145 \ldots 150$~MeV, the maximum of scaled interaction measure at $\tc ^{(e-3p)T^4}=205 \ldots 210$~MeV, as well as the inflection points of scaled pressure at $\tc^{p/T^4}=180\ldots190$~MeV and scaled entropy density at $\tc^{s/T^3}=155\ldots170$~MeV according to the tables in \cite{bor14,baza14}. Often one quotes $T_{\rm pc} = (155 \pm 9)$~MeV as pseudo critical temperature determined by the chiral susceptility, for instance, cf.~\cite{baza14,ess1, ess2}. The pseudo critical curve is parametrized as $\tc(\mu)=\tc(0) (1+\kappa \mu^2+\ldots)$ with displaying $\tc(0)=$145, 155 and 165~MeV, e.g.~in \cite{ess3}. This is to be complemented by the chemical freeze-out temperature $T_{\rm f.o.}=156$ MeV \cite{ess4, Stachel3} or $164 \ldots 168$~MeV \cite{ess5} not to be mixed with the kinetic freeze-out temperature; other recent claims are $T_{\rm f.o.} =165$~MeV (with afterburner) vs.~$T_{\rm f.o.} =155$~MeV (without afterburner) \cite{ess6}. \cite{bor14} mentions also a hadronization temperature of $164$~MeV. Here, we identify $\tc^{c_s^2}=\tc$.} \\
Alternatively, one can use the relation
  \begin{equation}
   z_H(T) = z_H^{(0)} \exp \li\{-\frac13 \int \limits_{T_0}^T \! \frac{\dd \tilde T}{\tilde T c_s^{2} (\tilde T)} \ri\} \label{zh}
  \end{equation}
to get the relation $T(z_H)$ from the lattice data $c_s^2(T)$ by exploiting the model input $\frac1{s} \frac{\dd s}{\dd z_H} = -3/z_H$ from (\ref{s}). The result is displayed in Fig.~\ref{c2_abb3} by the gray band which arises from using the lower and upper error bars of the data. In addition, the curves of Fig.~\ref{c2_abb2} are displayed too, proving that the ansatz (\ref{T}) is in fact meaningful, supposed (\ref{s}) holds true. A key ingredient here is the ansatz (\ref{s}) which is as in the extended soft wall model, \cite{Col12}, while (\ref{T}) is already beyond the model \cite{Col12} to lift the melting temperature of vector mesons, $T_{\rm dis}$, in the order of $\tc$.

\section{Melting of the vector mesons}
\begin{figure}
   \cen{\includegraphics[scale=.7]{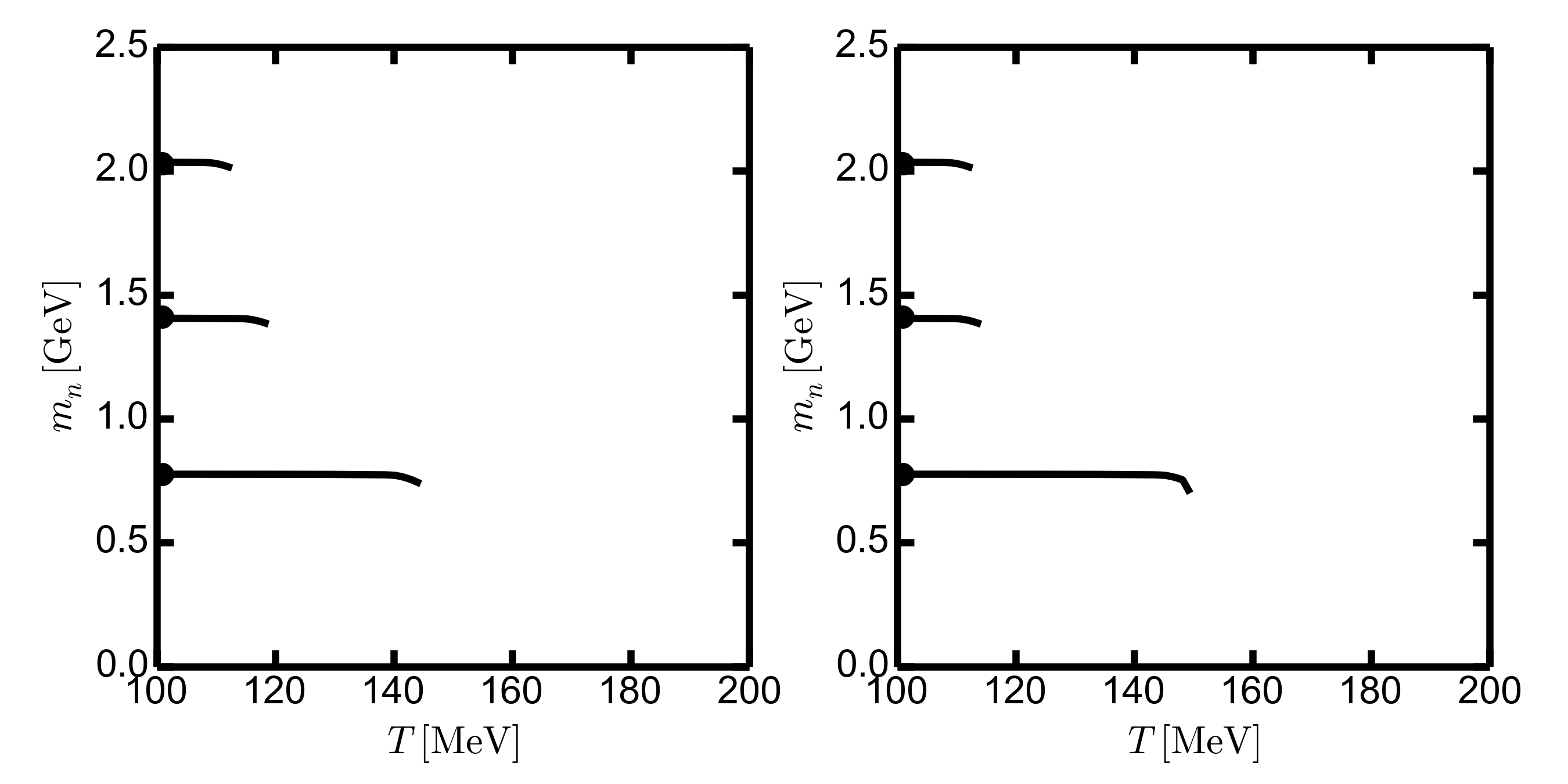}
   \caption{Vector meson masses $m_n$ for $n=0$ ($\rho(770)$), $n=1$ ($\rho(1450)$) and $n=2$ ($\rho(2030)$) as a function of the temperature. Left (right) panel: parameter set $\{T_x,\theta,b\}$ as for the dashed curves in Fig.~\ref{c2_abb2}-left (right) with the additional parameters $p=4$ and $c=310$~MeV, which cause vacuum masses as indicated by the bullets.}\label{c2_abbC}}
  \end{figure}

There is a quantitative tension of the above thermodynamics and the melting of vector mesons. The extended soft wall model (see \ref{appa} for a reminder) requires specific parameters $T_x$, $\theta$ and $b$ to ensure that vector mesons exist at temperatures $T\leq T_{\rm dis} = \mathcal{O}(150 \, {\rm MeV})$. The latter value comes from the thermo-statistical model analyses \cite{Stachel2,Stachel3, CKWX,Bec2, thst1,thst3} of hadron yields in ultra-relativistic heavy-ion collisions and is supported by analyses of fluctuation measures \cite{fl1,fl2,fl3,fl4,fl5}. Above $T_{\rm dis}$, hadrons are `molten`, at least these ones built up by light quarks, such as the $\rho$ mesons and its first excitations. At $T_{\rm dis}$ hadrons form and persist at temperatures below $T_{\rm dis}$. To calculate this temperature dependence one has to fix additionally $A(z)$ and $f(z)$. We employ here
  \begin{eqnarray}
   A(z) &=& -2 \ln (z/L), \, \, Lc=1, \label{A} \\
   f(x=z/z_H;z_HT(z_H))&  = &1-x^4  \li\{1+\frac{\ln \hat \varphi}{e \hat \varphi}\li[x^{2\hat\varphi}-1\ri]\ri\}, \label{f} \\
   \hat \varphi(z_HT(z_H)) &=& \exp\li\{\frac2e (\pi z_H T(z_H)-1)\ri\}
  \end{eqnarray}
which ensure $s(z_H) = \exp\{3/2 A(z_H) \}\propto {L^3}/{z_H^3} $ and $T(z_H)=-(\Dp_z f)\! \!\mid _{z_H} /4\pi$ as well as the proper asymptotic AdS behavior. Note that the dimensional scale parameter $c$ in the dilaton profile enters the calculation of $T_{\rm dis}^{\rho}$ (see \ref{appa}). \\
The $\chi^2$ minimization for the sound velocity squared (see Fig.~\ref{c2_abb2}, solid curves) constrains the parameters $T_x$, $\theta$ and $b$ such to deliver $T_{\rm dis}^{\rho} \approx 110$ MeV, while the adjustment to the softest point (see Fig.~\ref{c2_abb2}, dashed curves) result in $T_{\rm dis}^{\rho}=145$~MeV, i.e.~quite close to the wanted value of $\mathcal{O}(150~{\rm MeV})$. However, even exploiting the variation of $c(p)$ (see Fig.~\ref{c2_abbB}-left) does not allow to reach exactly that value. \\
Another way to visualize such a tension is to exhibit in Fig.~\ref{c2_abb3} a curve $T(z_H)$ for parameters which give $T_{\rm dis}^{\rho} = 155$~MeV (see dotted curve): This curve deviates significantly from the gray band in the low-temperature regime.\\
A remaining issue is whether the ground-state vector meson mass near and slightly below $\tc$ suffer strong medium modifications. Figure \ref{c2_abbC} shows in fact that, despite of a minor temperature dependence at the melting point, the vacuum mass is recovered in our model. This is remarkably consistent with the thermo-statistical model analyses of data \cite{Stachel2,Stachel3, CKWX,Bec2, thst1,thst3}. We emphasize that the first and the second excitation appears only at significantly lower temperature, and higher excitations are not found a $T>100$~MeV. (For a parameter selection, where a few first excitations on a Regge trajectory extend to $\tc$, cf.~\cite{ich2}). Note that the effect of an increasing value of $p$ implying an increasing altitude of the r.h.s.~potential wall prevails over the influence on $\tc$ of a decreasing scale parameter $c$ up to $p\approx 4$. Thus by the variation of $p$ one is able to lift $T_{\rm dis}$ a little bit. \ref{appb} shows a more efficient manner to lift $T_{\rm dis}$.

\section{Summary}
In summary we employ the extended soft wall model and show the compatibility of some useful optional model parametrizations with QCD thermodynamics near $\tc$. The quantity $\tc$ is here understood as the temperature of the softest point (minimum of sound velocity) and thus related to an observable. Within the hitherto employed model ans\"atze, the QCD lattice data can be perfectly reproduced. At the same time, the such fixed parameters propagate into another branch of the soft wall model where the temperature dependence of light-quark vector meson states is evaluated. Keeping the ans\"atze for the AdS-black hole metric functions and the dilaton profile as in previous work, we find that the $\rho$ meson melting at the temperature of $T_{\rm dis}^{\rho}=\tc$ is still compatible with lattice QCD thermodynamics. It is the low value of $\tc \cong 145$~MeV which points to a slight quantitative tension with our imagination of a hadronization temperature of $T_{\rm dis}^{\rho} \approx 150 \ldots 155$~MeV (or slightly above) to apply the thermo-statistical model for hadrons with vacuum masses. Given the simple structures of our ans\"atze it seems not too surprising that the available parameter space is rapidly exhausted in accommodating such vastly different aspects as the thermodynamics of the strongly interacting medium and the fate of selected individual constituents thereof. This calls for a self consistent holographic model where the profiles of the metric functions and the dilaton as conformal symmetry breaker are solutions of the corresponding gravity dual. 

\appendix 

\section{Recalling the extended soft wall model} \label{appa}
In the probe limit and focusing on a vector-like gauge field $V_M$ the action of the soft wall model reads \cite{KKSS}
  \begin{equation}
   S_V = \frac1k  \int  \! \dd^4 x\, \dd z \,   \sqrt{g} e^{-\Phi(z)} F^2,  \label{wirkung}
  \end{equation}
where $F^2 = g^{MM'}g^{NN'} F_{MN}F_{M'N'}$ with $F_{MN} = \partial_M V_N - \partial_N V_M$ ($M,N = 0, \ldots, 4$) denoting the field strength tensor of the $U(1)$ vector field $V$, where the gauges $V_4=0$ and $\partial^\mu V_\mu = 0$ can be applied by Klein-Kaluza decomposition. $V_{\mu}$ is then sourced by a current operator $\bar q \gamma_{\mu} q$ in the boundary theory in the spirit of the field-operator correspondence; the determinant of the metric $g_{MN}$ is again $g$, and $k$ stands for the gauge coupling with the AdS radius $L$ but drops out in the sequel. Greek indices run in the range $0 \ldots 3$. The vector field $V_{\mu}= \epsilon_\mu \varphi(z) \exp( i p_{\nu}  x^{\nu} )$ has the polarization described by $\epsilon_{\mu}$; the important part of the wave function is the amplitude $\varphi(z)$ beyond the phase $p_{\nu}x^{\nu}$. 

  \begin{figure}
   \cen{\includegraphics[scale=.7]{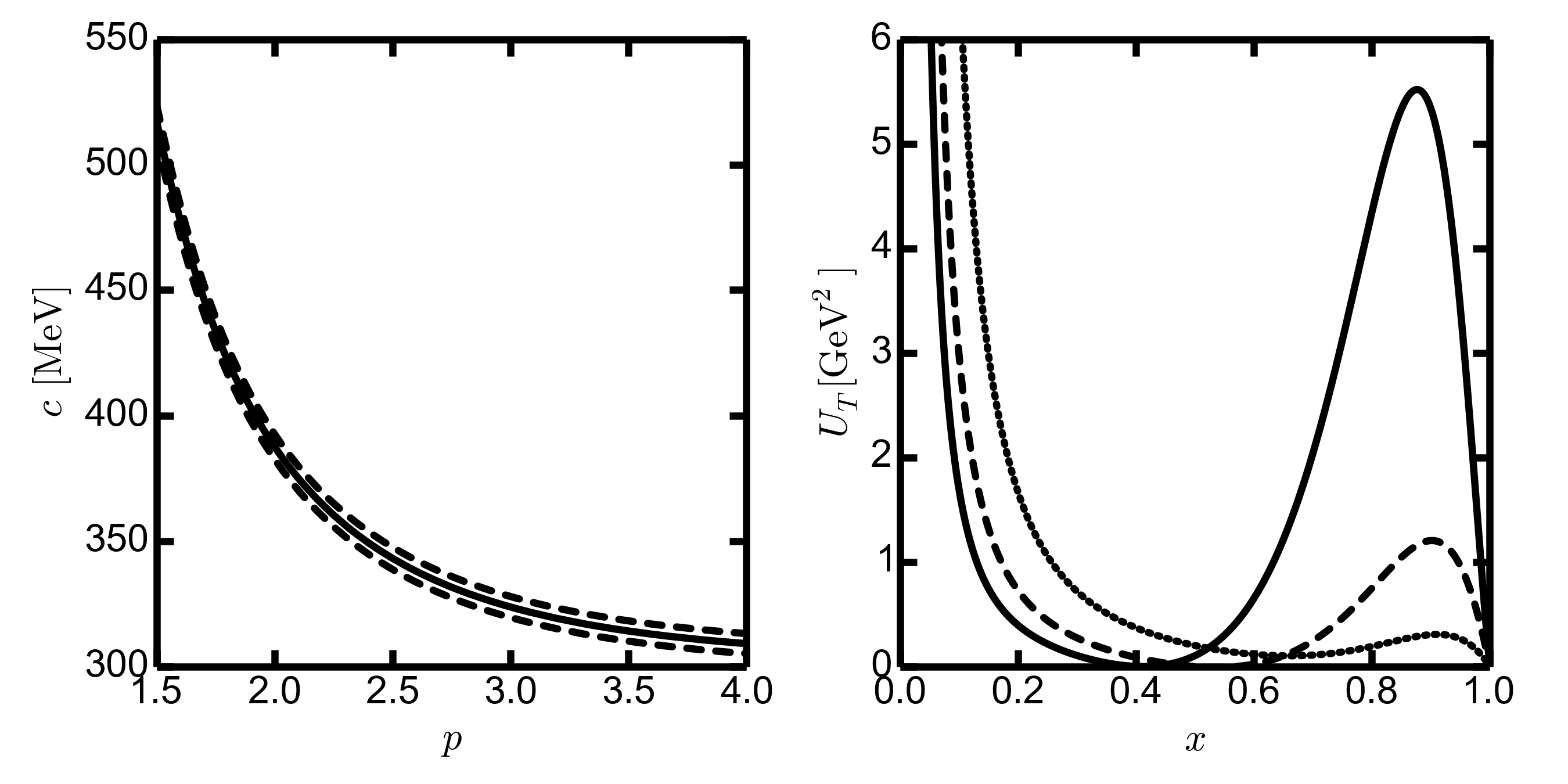}
   \caption{Left panel: The dependence $c(p)$ arising from the adjustment $m_0=m_{\rho}^{\rm (exp)}=770$~MeV with $\pm 10$~MeV variation. Right panel: The Schr\"odinger equivalent potential $U_T$ (\ref{hotpot}) as a function of $x=z/z_H$ for the parameter set $T_x=130$ MeV, $\theta=1.8$, $b=0.7$, $c=310$ MeV and $p=4$ for various temperatures ($T=105$ MeV (solid), $T=130$ MeV (dashed) and $T=150$~MeV (dotted)). Note that for solving (\ref{schroe}), $U_T$ as a function of the tortoise coordinate $\xi(z)$ is needed.} \label{c2_abbB}}
  \end{figure}
  
The resulting equation of motion with metric (\ref{ds}) can be cast into the form of a one-dimensional Schr\"odinger equation
  \begin{equation}
   \li(\Dp_{\xi}^2 -(U_T-m_n^2) \ri) \psi=0 \label{schroe}
  \end{equation}
by a coordinate transformation $z \to \xi$ with $\dd \xi (z) / \dd z= 1 /f(z)$ and the transformation $\psi =  \varphi \, \exp\{-\frac{1}{2}(A - \Phi)\} $. The potential reads
  \begin{equation}
   U_T = \li(\frac12 (\frac12 \Dp_{z}^2 A-\Dp_{z}^2 \Phi) +\frac14 (\frac12 \Dp_{z} A-\Dp_{z} \Phi)^2 \ri) f^2+ \frac14 (\frac12 \Dp_{z} A-\Dp_{z} \Phi) \Dp_{z} f^2. \label{hotpot}
  \end{equation}
The vector meson states correspond to normalizable solutions of (\ref{schroe}) with masses squared $m_n^2=p_{\mu}p^{\mu}$. The case $T=0$ is equivalent to setting $f=1$. The remaining warp factor $A(z)$ and the dilaton profile $\Phi(z)$ are to be adjusted suitably. In line with the original soft wall model \cite{KKSS}, an easy ansatz is (\ref{A}) and $\Phi(z)=(cz)^p$. We obtain the dependence $c(p)$ by the requirement $m_0=m_{\rho}^{\rm (exp)}=770$~MeV, see Fig.~\ref{c2_abbB}-left panel. 
In such a way, the hadron energy scale $c$ is fixed. At non-zero temperature, we use the blackness function (\ref{f}) and keep the above profiles $A(z)$ and $\Phi(z)$. An example of the potential $U_T$ (\ref{hotpot}) as a function of the scaled bulk coordinate $x=z/z_H$ is exhibited in Fig.~\ref{c2_abbB}-right. With increasing temperature the r.h.s.~potential wall diminishes rapidly so that no bound state can be accommodated at $T>T_{\rm dis}$. The temperature dependence of the lowest vector meson masses is exhibited in Fig.~\ref{c2_abbC}.

\section{Modified warp factor} \label{appb}
Solving the field equations emerging from (\ref{dilatonwirkung}) points to a more involved profile of the warp function $A(z)$ than that given in (\ref{A}). This in turn modifies the entropy (\ref{s}). Let us consider, instead of (\ref{s}), the trial ansatz
	\begin{equation}
	s(z_H) = 16\pi G_5 \frac{e^{\ve z_H}}{z_H^3} \label{sneu}
	\end{equation}
	
   \begin{figure}
   \cen{\includegraphics[scale=.7]{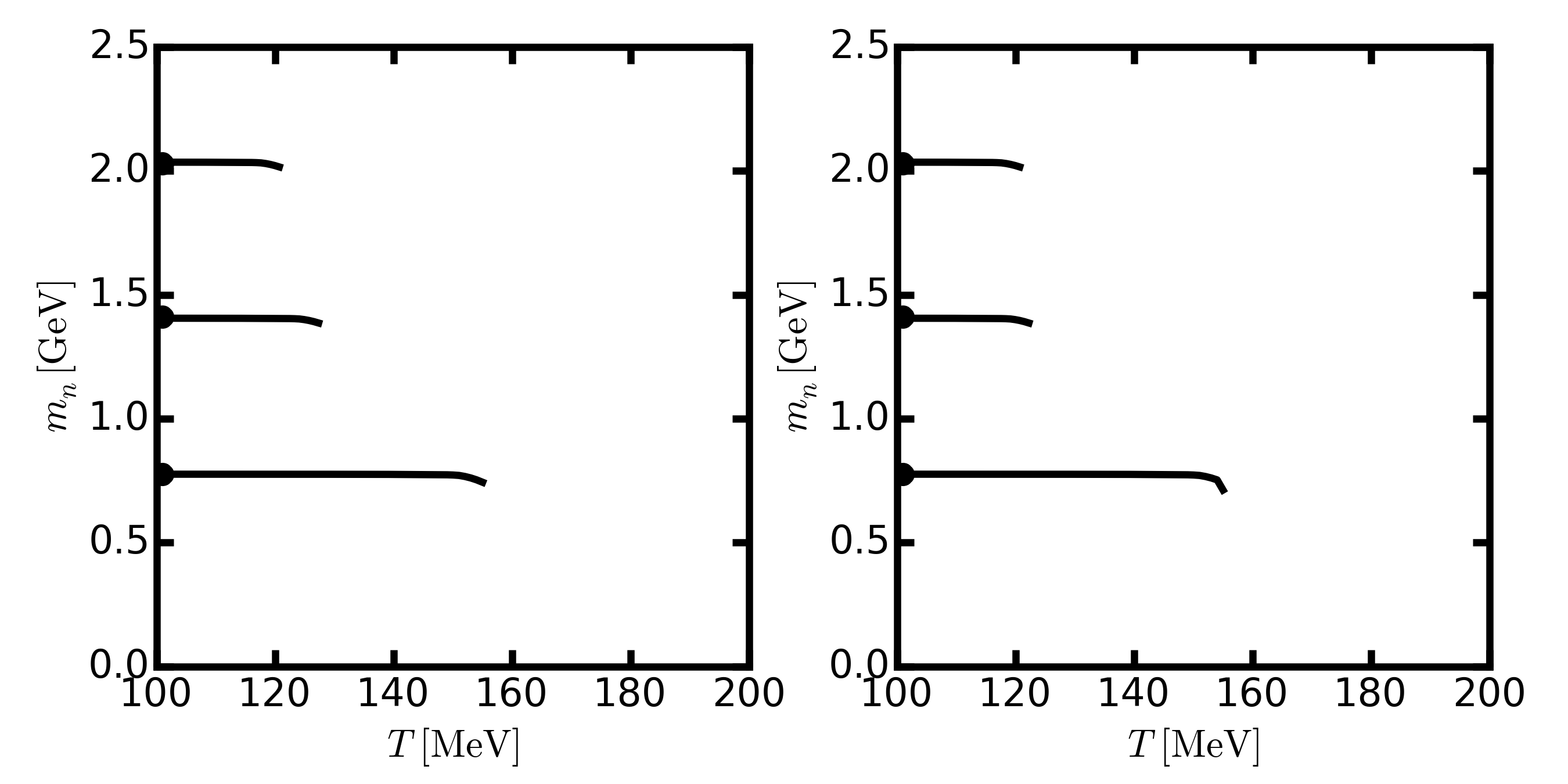}
   \caption{As in Fig.~\ref{c2_abbC}, but now for the entropy (\ref{sneu}); $c=310$~MeV and $p=4$. For further parameters cf.~Fig.~\ref{c2_abbA}.}\label{c2_abbD}}
  \end{figure}

  \begin{figure}[ht]
   \cen{\includegraphics[scale=.7]{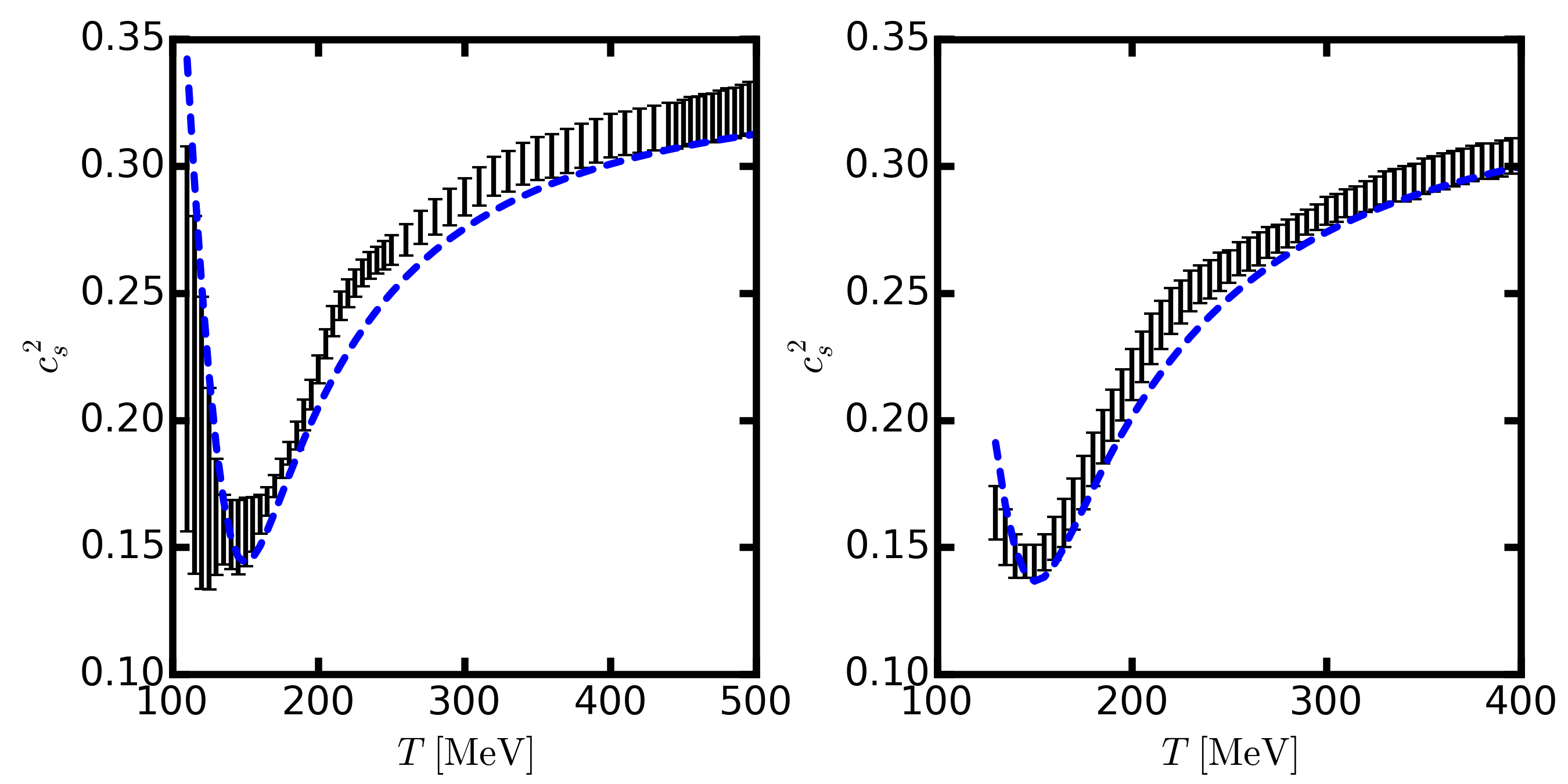}
   \caption{As in Fig.~\ref{c2_abb2}, but for the entropy (\ref{sneu}). Parameters are $T_x=$134 (136) MeV, $\theta=$1.98 (2.04), $b=$0.67 (0.68) and $\ve=$12.3 (17.4) MeV for the left (right) panel and $\tc=150$~MeV.}\label{c2_abbA}}
  \end{figure}

with the small positive dimensional parameter $\ve$ which results from an additional polynomial term in $A(z)$. The impact on the Schr\"odinger equivalent potential $U_T$ (\ref{hotpot}) is very tiny. However, (\ref{sneu}) allows to select values of $b$, $\theta$ and $T_x$ such that the disappearance temperature $T_{\rm dis}^{\rho}$ can be tuned to 155~MeV (see Fig.~\ref{c2_abbD}), while the consistency with the speed of sound data is preserved (see Fig.~\ref{c2_abbA}).

\ack{The work is supported by Studienstiftung des deutschen Volkes.}

\vspace{0.5cm}
 
\hrule

\end{document}